\documentstyle[aps]{revtex}

\begin{document}
\author{E. J. Ferrer$^{1},$ V. de la Incera$^{1},$ and A. Romeo$^{2}$}
\address{$^{1}$Department of Physics, State University of New York at Fredonia,\\
Houghton Hall 118, Fredonia, NY 14063, USA\\
$^{2}$Institute for Space Studies of Catalonia, CSIC, Edif. Nexus, Gran\\
Capita 2-4, 08034 Barcelona, Spain.}
\title{Spontaneous CPT Violation in Confined QED}
\maketitle

\begin{abstract}
Symmetry breaking induced by untwisted fermions in QED in a nonsimply
connected spacetime with topology $S^{1}\times R^{3}$ is investigated. It is
found that the discrete CPT symmetry of the theory is spontaneously broken
by the appearance of a constant vacuum expectation value of the
electromagnetic potential along the direction of space periodicity. The
constant potential is shown to be gauge nonequivalent to zero in the
nonsimply connected spacetime under consideration. Due to the symmetry
breaking, one of the electromagnetic modes of propagation is massive with a
mass that depends on the inverse of the compactification length. As a
result, the system exhibits a sort of topological directional
superconductivity.
\end{abstract}

\vskip5mmSymmetry breaking in non-trivial spacetime, where the curvature, as
well as the topology, can play an important role, has received much
attention in the last years because of its possible relevance for cosmology
in the early universe.

It is well known that the global properties of the spacetime, even if it is
locally flat, can give rise to new physics. The seminal discovery in this
direction is the so called Casimir effect \cite{Casimir}$,$ that is, the
existence of an attractive force between neutral parallel perfectly
conducting plates. In this phenomenon the attractive force appearing between
the plates is mediated by the zero-point fluctuations of the electromagnetic
field in vacuum. Thus, the Casimir force is interpreted as a macroscopic
manifestation of the vacuum structure of the quantized fields in the
presence of domains restricted by boundaries or nontrivial topologies\cite
{Milton}$.$

A characteristic of quantum field theory in spacetime with non-trivial
topology is the possible existence of nonequivalent types of fields with the
same spin\cite{Isham}$.$ In particular, for a fermion system in a space-time
which is locally flat but with topology represented by the domain $%
S^{1}\times R^{3}$ (i.e. a Minkowskian space with one of the spatial
dimensions compactified in a circle $S^{1}$ of finite length $a$), the
non-trivial topology is transferred into periodic boundary conditions for
untwisted fermions or antiperiodic boundary conditions for twisted fermions

\begin{equation}
\psi \left( t,x,y,z-a/2\right) =\pm \psi \left( t,x,y,z+a/2\right)  \label{1}
\end{equation}
In (\ref{1}) the compactified dimension with length $a$ has been taken along
the ${\cal OZ}$-direction.

Quantum electrodynamics (QED) with photons coupled to untwisted fermions or
to a combination of twisted and untwisted fermions is an unstable theory\cite
{Ford}$,$\cite{Ford2}. The instability arises due to polarization effects of
untwisted electrons that change the propagation of electromagnetic waves
giving rise to tachyonic electromagnetic modes\cite{Ford}$,$\cite{Ford2}$.$
In Ref. \cite{Ford2} it was speculated that such instability might indicate
that a stable vacuum configuration with nonzero expectation value of the
strength tensor $F_{\mu \nu }$ could appear.

For self-interacting scalar fields the space periodicity can also produce
instabilities causing a massless field to become massive\cite{Toshimura}$,$%
\cite{Toms1}$.$ The acquired mass depends on the periodicity length, so the
phenomenon is called topological mass generation.

Our main goal in this paper is to investigate the symmetry breaking induced
by the non-trivial topology $S^{1}\times R^{3}$ in the case of QED with
untwisted fermions. We will show that the stable vacuum configuration is
given by a constant electromagnetic potential $A_{3}$ along the compactified
dimension. As discussed below, even though such a vacuum configuration has $%
F_{\mu \nu }=0$, it cannot be gauged to zero, because the gauge
transformation that would be needed does not respect the periodicity of the
function space in the $S^{1}\times R^{3}$ domain. This is a sort of
Aharonov-Bohm effect which takes place due to the non-simply connected
topological structure of the considered spacetime. The nonzero vev of the
electromagnetic potential spontaneously breaks the discrete CPT symmetry of
the theory. Accordingly, the $A_{3}$ field acquires a topological mass and
no tachyonic mode remains. The results here reported can be of interest for
theories with extra dimensions on which some of them are compactified, and
for condensed matter quasi-planar systems.

The existence of constant gauge potentials with physical meaning are known
in statistics, where $A_{0}$ cannot be gauged away by reasons similar to
those pointed out above\cite{Batakis}$.$ The physical relevance of constant
condensates of zero-components of gauge fields has been investigated in the
context of many-particle electroweak theory in Refs. \cite{Linde}$,$\cite
{Incera}$.$

Let us consider the QED action in a spacetime domain with compactified
dimension of length $a$ in the ${\cal OZ}$-direction

\begin{equation}
{\cal S}=\int\limits_{-a/2}^{a/2}dx_{3}\int\limits_{-\infty }^{\infty
}dx_{0}d^{2}x_{\bot }\left[ -\frac{1}{4}F_{\mu \nu }^{2}+\overline{\psi }%
(i\partial \llap / -eA\llap / -m)\psi \right] ,\qquad  \label{2}
\end{equation}

The action (\ref{2}), together with the boundary conditions (\ref{1}),
represents a system of photons and electrons confined between two infinite
parallel planes perpendicular to the ${\cal OZ}$-direction.

When this confined QED action is considered for untwisted fermions, the
effect of vacuum polarization upon photon propagation yields a tachyonic
mass for the third component of the photon field \cite{Ford}. Tachyonic
modes in a quantum theory are an indication that the vacuum under
consideration is not really the physical one. A tachyonic mode is therefore
related to some symmetry breaking mechanism which leads to the true physical
vacuum. It follows that some symmetry breaking should then occur in the
unstable theory (2) of photons interacting with untwisted fermions.

Henceforth we restrict our analysis to this unstable case. To find the
physical vacuum that stabilizes the untwisted fermion theory, we propose the
following ansatz\footnote{%
This is a natural ansatz given that the ${\cal OZ}$-direction is the only
distinguished direction in the $S^{1}\times R^{3}$ space under
consideration, and that the $A_{\mu }$ field obeys periodic boundary
conditions along the ${\cal OZ}$-direction.} 
\begin{equation}
\overline{A}_{\nu }=\Lambda \delta _{\nu 3}  \label{20}
\end{equation}
for the vacuum solution, with $\Lambda $ an arbitrary constant that will be
determined from the minimum equation of the effective potential. Notice that
a constant electromagnetic potential in the direction of the periodicity ($%
\nu =3$ in our case), although seemingly physically equivalent to the pure
vacuum $\overline{A}_{\nu }=0$, since both have $F_{\mu \nu }=0$, cannot be
gauged to zero. The reason is that due to the periodicity of the fields in
the $S^{1}\times R^{3}$ space, the gauge transformations $A_{\mu
}\rightarrow A_{\mu }-\frac{1}{e}\partial _{\mu }\alpha $ are restricted to
those satisfying $\alpha (x_{3}+a)=$ $\alpha (x_{3})+2l\pi ,$ $l\in {\em Z}$%
\cite{Batakis}$.$ However, the gauge transformation $\alpha (x)=(x\cdot
n)e\Lambda $ connecting the constant field configuration (\ref{20}) with
zero does not satisfy the required periodicity condition unless $\Lambda $
satisfies

\begin{equation}
\Lambda =\frac{2l\pi }{ea},\qquad l\in {\em Z}  \label{200}
\end{equation}

Let us consider then the one-loop effective potential of the theory (\ref{2}%
) around the vacuum configuration (\ref{20})

\begin{equation}
V=-\frac{1}{2}a^{-1}\ln \left( Det\,\overline{G}^{-1}\right)  \label{20a}
\end{equation}
Here $Det\,\overline{G}^{-1}=$ $%
\sum\hspace{-0.4cm}\int%
d^{4}p\det \,\overline{G}^{-1}=\sum\limits_{p_{3}}\int d^{3}p\det \,%
\overline{G}^{-1},$ with $p_{3}=2n\pi /a,$ $(n=0,\pm 1,\pm 2,...)$ being the
discrete frequencies associated with periodic fermions. The fermion inverse
Green's function in the background $\Lambda $ is

\begin{equation}
\overline{G}^{-1}=\gamma \cdot \overline{p}+m  \label{20aa}
\end{equation}
with $\overline{p}_{\mu }=\left( p_{0},{\bf p}_{\perp },p_{3}-e\Lambda
\right) $. After summing in $p_{3}$ we obtain

\begin{equation}
V(\Lambda )=-\int\limits_{-\infty }^{\infty }\frac{d^{3}\widehat{p}}{\left(
2\pi \right) ^{3}}\left[ \varepsilon _{p}+2a^{-1}%
\mathop{\rm Re}%
\ln \left( 1-e^{-a\left( \varepsilon _{p}-ie\Lambda \right) }\right) \right]
.  \label{20b}
\end{equation}
\noindent

The extremum of the effective potential (\ref{20b}) satisfies

\begin{equation}
\left. \frac{\partial V(\Lambda )}{\partial \Lambda }\right| _{\Lambda
=\Lambda _{\min }}=-\int\limits_{-\infty }^{\infty }\frac{d^{3}\widehat{p}}{%
\left( 2\pi \right) ^{3}}\frac{2e^{-a\varepsilon _{p}}\sin \left( ae\Lambda
\right) }{1+e^{-2a\varepsilon _{p}}-2e^{-a\varepsilon _{p}}\cos \left(
ae\Lambda \right) }=0,  \label{25a}
\end{equation}

The solution to (\ref{25a}) is $\Lambda =\frac{l\pi }{ea}$, $l\in {\em Z}$.
Nevertheless, the minimum condition $\partial ^{2}V(\Lambda _{\min
})/\partial ^{2}\Lambda >0$ is only satisfied by the subset

\begin{equation}
\Lambda _{\min }=\frac{\left( 2l+1\right) \pi }{ea},\qquad l\in {\em Z}
\label{26a}
\end{equation}

The elements in the set of minimum solutions (\ref{26a}), are gauge
equivalent since they are all connected by allowed gauge transformations ($%
\alpha (x_{3}+a)=$ $\alpha (x_{3})+2l\pi $). It should be pointed out,
however, that the solutions (\ref{26a}) are gauge nonequivalent to the
trivial vacuum $\Lambda =0,$ since none of them satisfies (\ref{200}).

Substituting with the minimum solution (\ref{26a}) in Eq. (\ref{20b}) we
obtain in the $am\ll 1$ approximation that the effective potential of
untwisted fermions reduces to

\begin{equation}
V(\Lambda _{\min })=-\frac{7\pi ^{2}}{360a^{4}}  \label{26aa}
\end{equation}
which coincides with the result reported for twisted fermions in Ref. \cite
{Ford}$.$ Thus, the vacuum energy of both classes of fermions coincides if
the corresponding correct vacuum solution is used.

As discussed above, the appearance of the constant vacuum solution $%
\overline{A}_{\nu }=\Lambda _{\min }\delta _{\nu 3}$ at the one-loop level
must be associated with symmetry breaking. It is easy to corroborate that if
the gauge field in the action (\ref{2}) is shifted by $A_{\mu
}(x)\rightarrow A_{\mu }(x)-\Lambda _{\min }\delta _{\mu 3}$, a new term $%
e\Lambda _{\min }\psi \gamma _{3}\overline{\psi }$ emerges violating the CPT
symmetry

\begin{equation}
x_{\mu }=-x_{\mu },\qquad A_{\mu }(x)=-A_{\mu }(-x),\qquad \psi (x)=-i\psi
(-x)\gamma _{5}\gamma _{0}  \label{26b}
\end{equation}
of the original action. We stress that the vacuum solution (\ref{26a})
breaks the CPT invariance by breaking each discrete symmetry separately,
while maintaining CP or any other product of two symmetries intact.

We conclude that the spatial compactification leads to spontaneous CPT
symmetry breaking with vacuum configuration given by a constant
electromagnetic potential whose amplitude increases with the decreasing of
the compactification radius as $1/a$.

Let us find the electromagnetic modes' masses in the background of the new
vacuum (\ref{26a}). With this end, the general structure of the
electromagnetic field Green's function in the $S^{1}\times R^{3}$ space
should be considered

\begin{equation}
\Delta _{\mu \nu }(k)=P(g_{\mu \nu }-\frac{k_{\mu }k_{\nu }}{k^{2}})+Q[\frac{%
k_{\mu }k_{\nu }}{k^{2}}-\frac{k_{\mu }n_{\nu }+n_{\mu }k_{\nu }}{(k\cdot n)}%
+\frac{k^{2}n_{\mu }n_{\nu }}{(k\cdot n)^{2}}]+\frac{\alpha }{k^{4}}k_{\mu
}k_{\nu },  \label{8}
\end{equation}
Here, due to the breaking of Lorentz invariance, in addition to the usual
tensor structures $k_{\mu }$ and $g_{\mu \nu }$, a spacelike unit vector $%
n^{\mu }=(0,0,0,1)$, pointing in the direction of the periodicity, must be
introduced. In Eq. (\ref{8}) $\alpha $ is a gauge fixing parameter
corresponding to the gauge condition $\frac{1}{\alpha }\partial _{\mu
}A_{\mu }=0$, and the coefficients $P$ and $Q$ are given by

\begin{equation}
P=\frac{1}{k^{2}+\Pi _{0}},\qquad Q=-\frac{\Pi _{1}}{(k^{2}+\Pi _{0})\left\{
k^{2}+\Pi _{0}-\Pi _{1}[k^{2}/(k\cdot n)^{2}+1]\right\} },  \label{9}
\end{equation}
The parameters $\Pi _{0}$, $\Pi _{1}$, are the coefficients of the
polarization operator $\Pi _{\mu \nu }$, whose general structure is

\begin{equation}
\Pi _{\mu \nu }(k)=\Pi _{0}(g_{\mu \nu }-\frac{k_{\mu }k_{\nu }}{k^{2}})+\Pi
_{1}[\frac{k_{\mu }k_{\nu }}{k^{2}}-\frac{k_{\mu }n_{\nu }+n_{\mu }k_{\nu }}{%
(k\cdot n)}+\frac{k^{2}n_{\mu }n_{\nu }}{(k\cdot n)^{2}}]  \label{10}
\end{equation}

The coefficient $\Pi _{1}$ is absent in the usual flat space case with
trivial topology. It arises here because of the explicit breaking of the
Lorentz invariance in the $S^{1}\times R^{3}$ spacetime. The situation is
similar to that in statistical quantum field theory\cite{Fradkyn}, where the
presence of a medium breaks the Lorentz invariance giving rise to the
compactification of the time variable. The role of $n_{\mu }$ in the finite
temperature case is played by the four-velocity of the medium $u_{\mu }$.

From (\ref{10}) it is easy to see that the polarization operator
coefficients can be expressed in terms of the two independent tensor
components $\Pi _{33}$ and $\Pi _{00}$ through the relations

\begin{equation}
\Pi _{0}=\frac{k^{2}}{\kappa }\left[ k_{3}^{2}k_{0}^{2}\Pi _{33}-\widehat{k}%
^{4}\Pi _{00}\right]  \label{11}
\end{equation}

\begin{equation}
\Pi _{1}=\frac{k^{2}k_{3}^{2}}{\kappa }\left[ {\bf k}^{2}\Pi _{33}-\widehat{k%
}^{2}\Pi _{00}\right]  \label{12}
\end{equation}
where we are using the notation $\widehat{k}_{\mu }\equiv (k_{0},{\bf k}%
_{\perp },0)$ and $\kappa =\widehat{k}^{2}({\bf k}^{2}\widehat{k}%
^{2}-k_{0}^{2}k_{3}^{2})$.

The electromagnetic modes' masses are found from the poles of the Green's
function (\ref{8}). From (\ref{9}), (\ref{11}) and (\ref{12}) we have that
for photons propagating in a direction perpendicular to the direction of
periodicity ($k_{3}=0$) the dispersion relations are

\begin{equation}
k_{0}^{2}-{\bf k}_{\perp }^{2}-\frac{\widehat{k}^{2}}{{\bf k}_{\bot }^{2}}%
\Pi _{00}=0  \label{13}
\end{equation}

\begin{equation}
k_{0}^{2}-{\bf k}_{\perp }^{2}-\Pi _{33}=0  \label{14}
\end{equation}
From Eqs. (\ref{13})-(\ref{14}) we see that the values of the $\Pi _{00}$
and $\Pi _{33}$ polarization operator components in the limit $k_{3}=0$, $%
\left| \widehat{k}\right| \rightarrow 0$ are crucial for the properties of
the electromagnetic field propagation in this non-trivial space, since they
play the role of the square masses of the electromagnetic modes ($M_{1}^{2}=(%
\widehat{k}^{2}/{\bf k}_{\bot }^{2})\Pi _{00}$, $M_{2}^{2}=\Pi _{33}$). The
appearance of the electromagnetic tachyonic mass for untwisted electrons
comes precisely from the fact that in the $a\left| \widehat{k}\right| \ll
am\ll 1$ limit, those polarization operator components, calculated on the
trivial electromagnetic vacuum ($\Lambda =0$), are given by

\begin{equation}
\Pi _{33}(k_{3}=0,\widehat{k})\simeq -\frac{8e^{2}}{a^{2}}\left[ \frac{1}{12}%
-\frac{\xi }{4}+{\cal O}(\xi ^{2})\right] +{\cal O}(\widehat{k}^{2})
\label{19}
\end{equation}

\begin{equation}
\Pi _{00}(k_{3}=0,\widehat{k})\simeq \frac{e^{2}}{3\pi ^{2}}{\bf k}_{\bot
}^{2}\left[ \frac{1}{4\xi }+\frac{1}{2}\ln \xi +{\cal O}(\xi ^{0})\right] +%
{\cal O}(k_{\bot }^{4})  \label{19a}
\end{equation}
where $\xi =am/2\pi \ll 1$ . The leading term in Eq. (\ref{19}) coincides
with the one reported in Ref. \cite{Ford}, after a factor correction%
\footnote{%
The result in Eq. (4.9) of Ref. \cite{Ford} should be multiplied by 2.}
previously pointed out in Ref. \cite{Toms2}. From (\ref{19}) and (\ref{19a})
we have that in the $k_{3}=0$,$\left| \widehat{k}\right| \rightarrow 0$
limit the electromagnetic modes' masses in the trivial vacuum are $%
M_{1}^{2}=0$ and $M_{2}^{2}<0$ (tachyonic mass) respectively.

To investigate the modification produced in the dispersion equations (\ref
{13})-(\ref{14}) by the nontrivial vacuum (\ref{26a}) the polarization
operator components $\Pi _{00}$, $\Pi _{33}$ must be calculated in the
background of the non-trivial vacuum state. With this aim, we start from the
free-fermion propagator for untwisted fermions in the non-trivial vacuum

\begin{equation}
G(x-x^{\prime })=\frac{1}{(2\pi )^{3}a}%
\sum\hspace{-0.5cm}\int%
d^{4}p\exp [ip\cdot (x-x^{\prime })]G(\widetilde{p})  \label{3}
\end{equation}
with

\begin{equation}
G(\widetilde{p})=\frac{\widetilde{p}\llap / -m}{\widetilde{p}%
^{2}-m^{2}+i\epsilon },\qquad \widetilde{p}_{\mu }=\left( p_{0},{\bf p}%
_{\perp },p_{3}-e\Lambda _{\min }\right)  \label{4}
\end{equation}

The polarization operator components corresponding to the Green's functions (%
\ref{3}) can be found to order-$e^{2}$ in the coupling constant from

\begin{equation}
\Pi _{\mu \nu }^{\Lambda }(k)=-\frac{4ie^{2}}{(2\pi )^{3}a}%
\sum\hspace{-0.5cm}\int%
d^{4}p\frac{\widetilde{p}_{\mu }(\widetilde{p}_{\nu }-k_{\nu })+\widetilde{p}%
_{\nu }(\widetilde{p}_{\mu }-k_{\mu })-\widetilde{p}\cdot (\widetilde{p}%
-k)g_{\mu \nu }+m^{2}g_{\mu \nu }}{(\widetilde{p}^{2}-m^{2})\left[ (%
\widetilde{p}-k)^{2}-m^{2}\right] }  \label{15}
\end{equation}

After summation in $p_{3}$ we find for the polarization operator components
entering in the dispersion relations (\ref{13})-(\ref{14})

\begin{equation}
\Pi _{33}^{\Lambda }(k_{3}=0,\widehat{k})=-\frac{e^{2}}{\pi ^{3}}\int d%
\widehat{p}\frac{2\varepsilon _{p}^{2}-\widehat{p}\cdot \widehat{k}}{%
\varepsilon _{p}\left( \varepsilon _{p-k}^{2}-\varepsilon _{p}^{2}\right) }%
n(p)  \label{18}
\end{equation}

\begin{equation}
\Pi _{00}^{\Lambda }(k_{3}=0,\widehat{k})=\frac{e^{2}}{\pi ^{3}}\int d%
\widehat{p}\frac{2p_{4}\left( p_{4}-k_{4}\right) +\widehat{p}\cdot \widehat{k%
}}{\varepsilon _{p}\left( \varepsilon _{p-k}^{2}-\varepsilon _{p}^{2}\right) 
}n(p)  \label{18a}
\end{equation}
where $\varepsilon _{p}=\sqrt{\widehat{p}^{2}+m^{2}}$, $\varepsilon _{p-k}=%
\sqrt{(\widehat{p}-\widehat{k})^{2}+m^{2}}$, and $n(p)=[1-\exp a(\varepsilon
_{p}\pm ie\Lambda _{\min })]^{-1}=(1+\exp a\varepsilon _{p})^{-1}$. Finally,
evaluating the integrals in (\ref{18}), (\ref{18a}) and considering the $%
a\left| \widehat{k}\right| \ll am\ll 1$ limit, we obtain 
\begin{equation}
\Pi _{33}^{\Lambda }(k_{3}=0,\widehat{k})\simeq \frac{e^{2}}{a^{2}}\left[ 
\frac{1}{3}+{\cal O}(\xi ^{2})\right] +{\cal O}(\widehat{k}^{2})  \label{28a}
\end{equation}

\begin{equation}
\Pi _{00}^{\Lambda }(k_{3}=0,\widehat{k})\simeq \frac{e^{2}}{3\pi ^{2}}%
k_{\bot }^{2}\left[ \frac{1}{2}\ln \xi +{\cal O}(\xi ^{0})\right] +{\cal O}%
(k_{\bot }^{4})  \label{28}
\end{equation}
It can be seen that (\ref{28a}) coincides with the result reported for
twisted fermions in trivial vacuum in Ref. \cite{Ford}. Using the results (%
\ref{28a}) and (\ref{28}) in the dispersion equations (\ref{13})-(\ref{14})
we find that the electromagnetic modes propagate on the nontrivial vacuum
with new masses $M_{1}^{2}\left[ \Lambda \right] =0$ and $M_{2}^{2}\left[
\Lambda \right] =e^{2}/3a^{2}>0$. It is interesting to notice that if we
make the change $a\rightarrow \beta $ ($\beta $ the inverse temperature) in $%
M_{2}^{2}\left[ \Lambda \right] $, then we would formally obtain the well
known result in statistical QED \cite{Fradkyn} of a Debye screening mass.

Then, we conclude that in the new vacuum state no tachyon is present.
Instead, a massive electromagnetic mode, with a mass that depends on the
inverse of the compactification radius $a$, arises. When the
compactification length is taken to infinity ($a\rightarrow \infty $), the
flat-space QED theory is regained with zero photon mass. Moreover, the
polarization of photons in the presence of twisted or untwisted fermions
will be equivalent at least up to the one-loop level. However, they take
place in different vacuums, the trivial one for twisted fermions and the
nontrivial $\overline{A}_{3}=\frac{\pi }{ea}$ for untwisted ones.

The existence of a massive electromagnetic mode in confined QED for both
untwisted and twisted fermions has also implications for the magnetic
response of the system. To understand this, let us consider the modified
Maxwell equation in linear response theory

\begin{equation}
\left[ \Box \delta _{\mu \nu }-\lambda _{M}^{-2}\delta _{3\mu }\delta _{3\nu
}\right] A_{\nu }=eJ_{\mu },  \label{29}
\end{equation}
taken in the Lorentz gauge $\partial _{\mu }A^{\mu }=0.$ In Eq. (\ref{29}),
we introduced the magnetic length $\lambda _{M}=1/M_{2}\left[ \Lambda
\right] $. Considering an external static and constant current $I\,$ flowing
inside the confined space along the ${\cal OZ}$-axis, the current density in
Eq. (\ref{29}) becomes $eJ_{3}(x)=I\,\delta ^{2}({\bf x}_{\bot })$. Then,
the induced potential, which is a solution of (\ref{29}) with periodic
boundary conditions in the ${\cal OZ}$-direction, will be

\begin{equation}
A_{3}(x)=-\frac{I}{2\pi }K_{0}\left[ \left| {\bf x}_{\bot }\right| /\lambda
_{M}\right] ,  \label{30}
\end{equation}
where $K_{0}$ denotes a modified Bessel function of the third kind. In the $%
\left| {\bf x}_{\bot }\right| \gg \lambda _{M}$ limit we obtain for the
corresponding magnetic field

\begin{equation}
{\bf B}(r)=\frac{I}{2\sqrt{2\pi }}\frac{\exp \left( -r/\lambda _{M}\right) }{%
\sqrt{r\lambda _{M}}}\widehat{{\bf \theta }}  \label{31}
\end{equation}
with $r=\left| {\bf x}_{\bot }\right| $ and $\widehat{{\bf \theta }}$
denoting the azimuth-angle unit vector in cylindrical coordinates. From (\ref
{31}) we see that an azimuthal magnetic field will be screened along the
radial direction, in a distance equal to the inverse of the topological mass 
$M_{2}\left[ \Lambda \right] $. When the separation between the two infinite
parallel plates decreases, the screening effect increases.

Our results indicate that confined QED exhibits a sort of topological
directional superconducting behavior, with a Meissner effect taking place
for magnetic fields induced by electric currents flowing in the direction of
the compactified dimension. This result could find applications in high-T$%
_{C}$ superconductivity where, as it is well known, the system is confined
to a quasi-two-dimensional space.

{\bf Acknowledgments }

It is a pleasure for two of the authors (EJF and VI) to express their
gratitude to the Institute for Spatial Studies of Catalonia and to the
University of Barcelona for the warm hospitality extended to them during the
time this work was completed. This work has been supported in part by NSF
grant PHY-9722059 (EJF and VI), NSF POWRE grant PHY-9973708 (VI).

\end{document}